\documentclass[aps,showpacs,preprintnumbers,amsmath,amssymb]{revtex4}
\UseRawInputEncoding
\usepackage{dcolumn}
\usepackage[dvips]{epsfig}

\usepackage{float}
\usepackage{graphics,epsfig}
\usepackage{graphicx}
\usepackage{epstopdf}
\usepackage{dcolumn}
\usepackage{bm}
\usepackage{gensymb}
\usepackage{footnote}
\usepackage{booktabs}
\usepackage{pdfpages}

\begin{document}
\baselineskip=0.8 cm
\title{Polarization distribution in the image of a synchrotron emitting ring around a regular black hole}

\author{Xueyao Liu$^{1}$,
Songbai Chen$^{1,2}$\footnote{Corresponding author: csb3752@hunnu.edu.cn},
Jiliang Jing$^{1,2}$ }
\affiliation{ $ ^1$ Department of Physics, Key Laboratory of Low Dimensional Quantum Structures
and Quantum Control of Ministry of Education, Synergetic Innovation Center for Quantum Effects and Applications, Hunan
Normal University, Changsha, Hunan 410081, People's Republic of China
\\
$ ^2$Center for Gravitation and Cosmology, College of Physical Science and Technology, Yangzhou University, Yangzhou 225009, People's Republic of China}

\begin{abstract}
\baselineskip=0.6 cm
\begin{center}
{\bf Abstract}
\end{center}

The polarized images of a synchrotron emitting ring are studied in the regular Hayward and Bardeen black hole spacetimes. These regular black holes carry a magnetic field in terms of gravity coupled to nonlinear electrodynamics. Results show that the main features of the polarization images of the emitting rings are similar in these two regular black hole spacetimes. As the magnetic charge parameter increase, the polarization intensity and the electric vector position angle in the image plane increase in Hayward and Bardeen black hole spacetimes. Moreover, the polarization intensity and electric vector position angle in the image of the emitting ring in the Hayward black hole spacetime are closer to those in the Schwarzschild case. The effects of the magnetic charge parameter on the Strokes $Q-U$ loops are also slightly smaller in the Hayward black hole spacetime. This information stored in the polarization images around Hayward and Bardeen black holes could help understand regular black holes and the gravity coupled to nonlinear electrodynamics.
\end{abstract}

 \maketitle
\newpage
\section{Introduction}

Observational black hole astronomy is widely considered to have entered an exciting era of rapid progress.
In recent years, the first image of the supermassive black hole M87* \cite{EHT1,EHT2,EHT3,EHT4,EHT5,EHT6} and its polarized image \cite{EHT7,EHT8} were released in succession by the Event Horizon Telescope collaboration.
The brightness of the surrounding emission region and the corresponding polarization patterns carry a wealth of information about the electromagnetic emissions near the black hole. Thus, studying the black hole image and its polarization patterns helps understand the matter distribution and accretion process around the black hole as well as its characteristics. In particular, analyzing polarization information in black hole images can further probe magnetic field configurations and powerful jets near a black hole, which is beneficial for obtaining insight into the physics in the strong-field region and checking theories of gravity. Thus, much effort has been devoted to investigating the polarized images of various black holes \cite{ts1,ts2,ts3,ts4,ts5,ts6,ts7,ts8,ts9,ts10,ts11,ts12,po2,ex}.

To obtain polarization information in black hole images in the observer's sky, null geodesic equations must be solved for photons traveling in the black hole spacetime together with the parallel transport equations of polarization vectors along photon geodesics. Generally, numerical simulations must be used to obtain an exact description of a polarized black hole image. Recently, a simple model has been used to investigate the polarized images of axisymmetric fluids orbiting black holes arising from synchrotron emission in various magnetic fields \cite{his1,po1}.
It is shown that the polarization signatures in the black hole images are dominated by the magnetic field configuration, together with the black hole parameters and the observer inclination. In this model, only the emission within a narrow range of radii $R$ is considered, but the image of a finite, thin disk can be produced by simply summing contributions from individual radii \cite{his1,po1, po3s}. With this model, the polarization signatures for a four-dimensional black hole in Gauss-Bonnet theory are analyzed in \cite{his2}.

In general relativity, according to the well-known singularity theorems \cite{Hawking}, the existence of singularities is inevitable, and black hole solutions own a singularity inside an event horizon. However, the singularities are widely believed to be nonphysical and are produced by classical theories of gravity and should be avoided in terms of the perfect theory of quantum gravity. In this spirit, the first regular black hole solution was proposed by Bardeen \cite{bm1} and is spherically symmetric without a singularity. However, the physical source of Bardeen black holes was unclear then. Until the end of the last century, a possible nonlinear electromagnetic source was proposed to account for Bardeen black holes \cite{hm2,hm2s}, so a regular black hole can be interpreted as the gravitational field of a nonlinear electric or magnetic monopole. Some other regular black hole solutions were obtained using the nonlinear electrodynamic mechanism \cite{hm1,me,me1,hm3}. The observational effects of regular black holes are of great research interest because they could help understand some fundamental issues in physics, including black holes, singularity and nonlinear electrodynamics \cite{nonlin1,nonlin2}. Since a regular black hole can be considered a solution of the gravity coupled to a nonlinear electromagnetic field, it should carry a magnetic field itself. This attribute could modify the polarized image of an emitting ring around a regular black hole. This paper aims to study the polarization information in the image of a synchrotron emitting ring around a regular Hayward black hole \cite{hm1} and a Bardeen black hole \cite{bm1} and to probe the effects of the magnetic charge parameter on the polarization image in these two regular black hole spacetimes.

The paper is organized as follows: Section II briefly introduces Hayward and Bardeen black holes and presents formulas to calculate the observed polarization vector in the image plane of an emitting ring in these two regular black hole spacetimes. Section III presents the polarization images of a synchrotron emitting ring around a regular black hole and probes the effects of the magnetic charge parameter on the polarization image. Finally, this paper ends with a summary.

\section{Observed polarization field in the regular black hole spacetimes}

This section focuses on regular Hayward and Bardeen black hole spacetimes. A Hayward black hole is an important static regular black hole and its metric has the form \cite{hm1}
\begin{equation}\label{metricline}
d s^{2}=-f(r) d t^{2}+\frac{1}{f(r)} d r^{2}+r^{2} d \theta^{2}+r^{2} \sin ^{2} \theta d \phi^{2},
\end{equation}
with
\begin{equation}\label{haywardff}
f(r)=1-\frac{2Mr^{2}}{r^{3}+g^{3}},
\end{equation}
where $M$ is related to the black hole mass, and $g$ is a regularizing parameter. The Hayward spacetime is asymptotically flat since the metric function $f\rightarrow 1$ as $r\rightarrow\infty$. Moreover, it is regular everywhere and contains no singularity. The Hayward spacetime can be considered a magnetic solution to the Einstein equation coupled to nonlinear electrodynamics with an action \cite{me,me1}
\begin{equation}\label{la1}
I=\frac{1}{16 \pi}\int d^{4} x \sqrt{-g}\left[ R-4 \mathcal{L}(\mathcal{F})\right].
\end{equation}
$R$ is the usual scalar curvature, and $\mathcal{L}(\mathcal{F})$ is the Lagrangian density with the form
\begin{equation}\label{la6}
\mathcal{L}(\mathcal{F})=\frac{12 }{\alpha} \frac{\left(\alpha \mathcal{F}\right)^{\frac{3}{2}}}{\left[1+\left(\alpha\mathcal{F}\right)^{\frac{3}{4}}\right]^{2}},
\end{equation}
where the invariant $\mathcal{F}\equiv\frac{1}{4}F^{\mu\nu}F_{\mu\nu}$ and the electromagnetic field tensor $F_{\mu\nu}=\partial_{\mu} A_{\nu}-\partial_{\nu} A_{\mu}$. $A_{\mu}$ is the electromagnetic four-vector potential. For the Hayward black hole spacetime, the electromagnetic four-vector potential
has the form
\begin{equation}\label{dcvect}
A_{\mu}=(0,0,0,Q_m\cos\theta),
\end{equation}
where $Q_{m}$ is the magnetic monopole charge. The coupling parameter $\alpha$ in the Lagrangian density (\ref{la6}) and $g$ in the metric function (\ref{haywardff}) are related to $Q_{m}$ by $\alpha=\frac{8Q_{m}^{6}}{M^{4}}$ and $g=\frac{2Q^2_{m}}{M}$, respectively. The metric function (\ref{haywardff}) can be rewritten as
\begin{equation}
f(r)=1-\frac{2M^{4}r^{2}}{8Q_{m}^{6}+M^{3} r^{3}}.
\end{equation}
Another important regular black hole is a Bardeen black hole \cite{bm1}. Its metric can be expressed as (\ref{metricline}), but the metric function is
\begin{equation}\label{bardeenff}
f(r)=1-\frac{2Mr^2}{(r^2+g^2)^{3/2}}.
\end{equation}
Similarly, $M$ is related to the black hole mass, and $g$ is a regularizing parameter. In \cite{me,me1,hm2,hm2s}, a Bardeen black hole is also considered a solution of the gravity coupled to nonlinear electrodynamics, and the corresponding Lagrangian density of the electromagnetic field is
\begin{equation}\label{labardeen}
\mathcal{L}(\mathcal{F})=\frac{12 }{\alpha} \frac{\left(\alpha \mathcal{F}\right)^{\frac{5}{4}}}{\left(1+\sqrt{\alpha\mathcal{F}}\right)^{\frac{5}{2}}}.
\end{equation}
With the magnetic monopole charge $Q_{m}$, the metric function $f(r)$ for a Bardeen black hole can be expressed as
\begin{equation}
f(r)=1-\frac{2M^{4}r^{2}}{(4Q_{m}^{4}+M^{2} r^{2})^{3/2}}.
\end{equation}
The electromagnetic four-vector potential $A_{\mu}$ has the same form in the Bardeen and Hayward black hole spacetimes. Thus, in these two regular black hole spacetimes, the nonzero components of the electromagnetic tensor are $F_{23}=-F_{32}=Q_{m}\sin\theta$. This result means that a regular black hole carries a magnetic field itself, which should modify the polarized image of the emitting ring around a black hole. As $Q_{m}=0$, these two regular black holes reduce to the Schwarzschild case.

The metric functions (\ref{haywardff}) and (\ref{bardeenff}) indicate that two regular black hole spacetimes coincide with the Schwarzschild solution for large $r$ and behave like the de Sitter spacetime for small $r$. Moreover, the scalar curvatures $R_{\mu\nu\rho\sigma}R^{\mu\nu\rho\sigma}$, $R_{\mu\nu}R^{\mu\nu}$, and $R$, respectively, can be expressed as
\begin{eqnarray}\label{RR1}
&&R_{\mu\nu\rho\sigma}R^{\mu\nu\rho\sigma}=\frac{48M^6(M^{12} r^{12}-32M^9Q^{6}_m r^9+1152M^6Q^{12}_m r^6-1024 M^3 Q^{18}_m r^3+8192 Q^{18}_m)}{(M^3r^3+8Q^6_{m})^6},\nonumber\\
&&R_{\mu\nu}R^{\mu\nu}=\frac{4608 M^8 Q^{12}_m (128 Q^{12} - 16 M^3 Q^6_m r^3 + 5 M^6 r^6)}{(8 Q^6_m + M^3 r^3)^6},\quad\quad R=\frac{96 M^4 Q^6_m (M^3 r^3-16 Q^6_m)}{(8 Q^6_m + M^3 r^3)^3)},
\end{eqnarray}
in the Hayward black hole spacetime and
\begin{eqnarray}\label{RR2}
&&R_{\mu\nu\rho\sigma}R^{\mu\nu\rho\sigma}=\frac{48M^6(M^8 r^8-12M^6Q^{4}_m r^6+188M^4Q^{8}_m r^4- 64 M^2 Q^{12}_m r^2+512 Q^{16}_m)}{(M^2r^2+4Q^4_{m})^7},\nonumber\\
&&R_{\mu\nu}R^{\mu\nu}=\frac{288 M^8 Q^8_m (128 Q^8_m- 16 M^2 Q^4_m r^2 + 13 M^4 r^4)}{(M^2r^2+4Q^4_{m})^7},\quad\quad R=\frac{24M^4 Q^4_m ( M^2 r^2-16 Q^4_m)}{(4Q^4_m+M^2 r^2)^{7/2}},
\end{eqnarray}
in the Bardeen black hole spacetime. These scalar curvatures are finite everywhere in two regular black hole spacetimes and differ from those in the Schwarzschild black hole spacetime with the divergent invariant $R_{\mu\nu\rho\sigma}R^{\mu\nu\rho\sigma}$ at the point $r=0$. In particular, each scalar curvature at $r=0$ has the same value in these two black holes. This attribute is understandable because these two spacetimes have the same behavior near the point $r=0$.
Moreover, the weak and strong energy conditions could be violated in the Hayward and the Bardeen black hole spacetimes \cite{weakenergy1,weakenergy2,weakenergy3,weakenergy4}. In contrast, the Schwarzschild black hole spacetime satisfies both these energy conditions. Notably, in the Bardeen (\ref{bardeenff}) or Hayward (\ref{haywardff}) spacetime, the two parameters $M$ and $g$ are not independent but are related to the coupling parameter $\alpha$ in the nonlinear electrodynamic theory, which is different from those in the Reissner-Nordstr\"{o}m case. This comparison implies that the Bardeen or Hayward solution could not be the most general static and spherically symmetric solution due to lacking an additional parameter associated with the condensate of the graviton. Thus, the current regularity in these two black hole spacetimes may be a fine-tuning result by setting this extra parameter to zero.

Now, the polarization vectors are to be studied for photons emitted from the ring around regular black holes.
A synchrotron emitting ring is assumed to lie in the equatorial plane of a regular black hole. In the local Cartesian frame of the point $P$ in the ring (the $P$-frame where the axis $\hat{x}$ is along the polar direction) \cite{his1,ex,po1,po2}, the nonzero components of the electromagnetic tensor become $F_{\hat{2}\hat{3}}=-F_{\hat{3}\hat{2}}=\frac{Q_{m}}{R^2}$, where $R$ is the ring radius. Then,
the black hole magnetic field in the $P$-frame can be written as
\begin{eqnarray}
\vec{B}_{(\mathrm{P})}=F_{\hat{2}\hat{3}}\hat{x}=\frac{Q_{m}}{R^2} \hat{x}.
\end{eqnarray}
Supposing that the fluid at point $P$ has a velocity $\beta$ with angle $\chi$ from the $\hat{x}$-axis in the local $P$-frame \cite{his1,ex,po1,po2}, i.e.,
\begin{equation}
\vec{\beta}=\beta(\cos \chi \hat{x}+\sin \chi \hat{y}),
\end{equation}
the magnetic field $B_{\mu(F)}$ and the photon's wave vector $k^{\mu}_{(F)}$ at point $P$ in the frame ($F$-frame) comoving with the fluid can be obtained through a Lorentz transformation \cite{his1,ex,po1,po2}
\begin{eqnarray}
B_{r(F)} = (\cos ^{2} \chi+ \gamma \sin ^{2} \chi) \frac{Q_{m}}{R^2}, \quad\quad\quad
B_{\phi(F)}= - \frac{(\gamma-1)Q_{m}}{R^2} \cos \chi \sin \chi ,
\end{eqnarray}
and
\begin{eqnarray}
k_{(F)}^{\hat{t}} & = & \gamma k_{(P)}^{\hat{t}}-\gamma \beta \cos \chi k_{(P)}^{\hat{x}}-\gamma \beta \sin \chi k_{(P)}^{\hat{y}}, \\\nonumber
k_{(F)}^{\hat{\hat{x}}} & = & -\gamma \beta \cos \chi k_{(P)}^{\hat{t}}+\left(1+(\gamma-1) \cos ^{2} \chi\right) k_{(P)}^{\hat{x}}+(\gamma-1) \cos \chi \sin \chi k_{(P)}^{\hat{y}}, \\\nonumber
k_{(F)}^{\hat{y}} & = & -\gamma \beta \sin \chi k_{(P)}^{\hat{t}}+(\gamma-1) \cos \chi \sin \chi k_{(P)}^{\hat{x}}+\left(1+(\gamma-1) \sin ^{2} \chi\right) k_{(P)}^{\hat{y}}, \\\nonumber
k_{(F)}^{\hat{z}} & = & k_{(P)}^{\hat{z}}.
\end{eqnarray}
where $\gamma$ is the Lorentz factor $ \gamma=\frac{1}{\sqrt{1-\beta^2}}$.
Setting $\zeta$ as the angle between the magnetic field $\vec{B}_{(F)}$ and the 3-vector $\vec{k}_{(F)}$ in the $F$-frame, the factor $\sin \zeta$ can be expressed as \cite{his1,ex,po1,po2}
\begin{equation}
\sin \zeta=\frac{\left|\vec{k}_{(\mathrm{F})} \times \vec{B}_{(F)}\right|}{\left|\vec{k}_{(\mathrm{F})}\right|\left|\vec{B}_{(F)}\right|},
\end{equation}
which plays an important role in the intensity of synchrotron radiation emitted along 3-vector $\vec{k}_{(F)}$. Since the electric vector of light is along the vector $\vec{k}_{(F)}\times \vec{B}_{(F)} $, the four-dimensional
polarization vector $f^{\mu}_{(F)}$ in the $F$-frame can be expressed as \cite{his1,ex,po1,po2}
\begin{equation}
f_{(\mathrm{F})}^{\hat{t}}=0, \quad f_{(\mathrm{F})}^{\hat{x}}=\frac{\left(\vec{k}_{(\mathrm{F})} \times \vec{B}_{(F)}\right)_{\hat{x}}}{\left|\vec{k}_{(\mathrm{F})}\right|}, \\\quad
f_{(\mathrm{F})}^{\hat{y}}=\frac{\left(\vec{k}_{(\mathrm{F})} \times \vec{B}_{(F)}\right)_{\hat{y}}}{\left|\vec{k}_{(\mathrm{F})}\right|}, \quad f_{(\mathrm{F})}^{\hat{z}}=\frac{\left(\vec{k}_{(\mathrm{F})} \times \vec{B}_{(F)}\right)_{\hat{z}}}{\left|\vec{k}_{(\mathrm{F})}\right|},
\end{equation}
which satisfies $f^{\mu} f_{\mu}=\sin ^{2} \zeta|\vec{B}_{(F)}|^{2}$. With the inverse Lorentz transformation \cite{his1,ex,po1,po2}, the polarization 4-vector $f^{\mu}_{(P)}$ in the $P$-frame can be given by
\begin{align}
f_{(\mathrm{P})}^{\hat{t}} & = \gamma f_{(\mathrm{F})}^{\hat{t}}+\gamma \beta \cos \chi f_{(\mathrm{F})}^{\hat{x}}+\gamma \beta \sin \chi f_{(\mathrm{F})}^{\hat{y}},\nonumber\\
f_{(\mathrm{P})}^{\hat{x}} & = \gamma \beta \cos \chi f_{(\mathrm{F})}^{\hat{\imath}}+\left(1+(\gamma-1) \cos ^{2} \chi\right) f_{(\mathrm{F})}^{\hat{x}}+(\gamma-1) \cos \chi \sin \chi f_{(\mathrm{F})}^{\hat{y}},\\
f_{(\mathrm{P})}^{\hat{y}} & = \gamma \beta \sin \chi f_{(\mathrm{F})}^{\hat{\imath}}+(\gamma-1) \cos \chi \sin \chi f_{(\mathrm{F})}^{\hat{x}}+\left(1+(\gamma-1) \sin ^{2} \chi\right) f_{(\mathrm{F})}^{\hat{y}},\nonumber\\
f_{(\mathrm{P})}^{\hat{z}} & = f_{(\mathrm{F})}^{\hat{z}} .\nonumber
\end{align}
Moreover, in regular black hole spacetimes $(\ref{metricline})$, the celestial coordinates $(x,y)$ for the photon
moving from point $P$ along the null geodesic to the observer at infinity are \cite{xy}
\begin{eqnarray}
&&x=-\frac{Rk^{\hat{y}}}{\sin\theta}, \\ \nonumber
&&y=R\sqrt{\left(k^{\hat{z}}\right)^2-\cot^2\theta\left(k^{\hat{y}}\right)^2} \operatorname{sgn}(\sin\phi).
\end{eqnarray}
With the help of the conserved Penrose-Walker constant $\kappa$ \cite{pc}, the polarization vector at the observer can be easily calculated because its real and imaginary parts are conserved along the null geodesic. At point $P$ in the fluid, the Penrose-Walker constant $\kappa$ has the form \cite{his1,ex,po1,po2}
\begin{eqnarray}\label{kappa0}
&&\kappa=\kappa_1+i\kappa_2, \quad\quad\quad
\kappa_1=\psi_{2}^{-1/3}\left(k^{t}f^{x}-k^{x}f^{t}\right), \quad\quad\quad \kappa_2=\psi_{2}^{-1/3}R^{2}\left(k^{y}f^{z}-k^{z}f^{y}\right),
\end{eqnarray}
with
\begin{eqnarray}
\begin{split}
&&k^{t}=\frac{1}{f(R)}, \quad\quad\quad\quad k^{x}=\sqrt{f(R)}k^{\hat{x}}_{(P)}, \quad\quad\quad\quad
k^{y}=\frac{k^{\hat{y}}_{(P)}}{R},\quad\quad\quad k^{z}=\frac{k^{\hat{z}}_{(P)}}{R}, \\
&&f^{t}=\frac{f^{\hat{t}}_{(P)}}{\sqrt{f(R)}}, \quad\quad\quad f^{x}=\sqrt{f(R)}f^{\hat{x}}_{(P)}, \quad\quad\quad\quad
f^{y}=\frac{f^{\hat{y}}_{(P)}}{R},\quad\quad\quad f^{z}=\frac{f^{\hat{z}}_{(P)}}{R}.
\end{split}
\end{eqnarray}
Here, the Weyl scalar $\psi_{2}$ has the form
\begin{equation}
\psi_{2}=\frac{M^{4}(M^{3} R^{3}-16 Q_{m}^{6})}{\left(8 Q_{m}^{6}+M^{3} R^{3}\right)^{2}},
\end{equation}
in the Hayward black hole spacetime and
\begin{equation}
\psi_{2}=\frac{M^{4}(M^{2} R^{2}-8  Q_{m}^{4})}{\left(4 Q_{m}^{4}+M^{2} R^{2}\right)^{5 / 2}},
\end{equation}
in the Bardeen black hole spacetime.
Thus, the normalized polarization electric field vector $\vec{E}$ along the $x$ and $y$ directions in the observer's sky can be given by \cite{his1,ex,po1,po2}
\begin{eqnarray}\label{EV0}
&&E_{x,norm}=\frac{y\kappa_2+x\kappa_1}{\sqrt{\left(\kappa_1^2+\kappa_2^2\right)\left(x^2+y^2\right)}},\nonumber \\
&&E_{y,norm}=\frac{y\kappa_1-x\kappa_2}{\sqrt{\left(\kappa_1^2+\kappa_2^2\right)\left(x^2+y^2\right)}}, \\
&&E_{x,norm}^2+E_{y,norm}^2=1.\nonumber
\end{eqnarray}
In general, for synchrotron radiation, the intensity of the linearly polarized light that reaches the observer from point $P$ can be approximated as \cite{his1,ex,po1,po2}
\begin{equation}\label{expression of intensity1}
|I|\sim\delta^{3+\alpha_\nu}l_p|\vec{B}|^{1+\alpha_\nu}\sin^{1+\alpha_\nu}\zeta,
\end{equation}
where the power $\alpha_\nu$ depends on the properties of the accretion disk, including the ratio of the emitted photon energy $h\nu$ to the disk temperature $kT$. The quantity $l_p$ is the geodesic path length for the photon traveling through the emitting region, which is given by \cite{his1,ex,po1,po2}
\begin{equation}
\quad l_p=\frac{k^{\hat{t}}_{(F)}}{k^{\hat{z}}_{(F)}}H.
\end{equation}
$H$ is the height of the disk and can be taken as a constant for simplicity. As in refs.\cite{his1,ex,po1,po2}, $\alpha_\nu$ can be set to $\alpha_\nu=1$, so the observed polarization intensity becomes
\begin{equation}
|I|=\delta^{4} l_{\mathrm{p}}|\vec{B}|^{2} \sin ^{2} \zeta,
\end{equation}
and the observed polarization vector components are
\begin{align}\label{EV01}
\begin{split}
E_{x, o b s} & = \delta^{2} l_{p}^{\frac{1}{2}} \sin \zeta|\vec{B}| E_{x, n o r m}, \\
E_{y, o b s} & = \delta^{2} l_{p}^{\frac{1}{2}} \sin \zeta|\vec{B}| E_{y, n o r m}.
\end{split}
\end{align}
Then, the total polarization intensity and the electric vector position angle (EVPA) can be expressed as \cite{his1,ex,po1,po2}
\begin{eqnarray}\label{Insen01}
I=E_{x,obs}^2+E_{y,obs}^2, \quad \quad EVPA=\frac{1}{2}\arctan\frac{U}{Q},
\end{eqnarray}
where the Stokes parameters $Q$ and $U$ are given by
\begin{eqnarray}\label{stroke01}
Q=E_{y,obs}^2-E_{x,obs}^2, \quad\quad\quad U=-2E_{x,obs}E_{y,obs}.
\end{eqnarray}
For regular black holes (\ref{metricline}), to obtain the polarization information in the image of point $P$, the null geodesic equation of the photon emitted from point $P$ must be solved first. Then, combining this solution with Eqs.(\ref{kappa0}), (\ref{EV0}), (\ref{EV01}), (\ref{Insen01}), and (\ref{stroke01}), the corresponding polarization intensity and EVPA in the pixel related to point $P$ can be obtained. Repeating similar operations along the ring, the total polarization image of the emitting ring around a regular black hole can be presented.

\section{Polarization images of the emitting ring around regular black holes}

\begin{figure}
\centering
\includegraphics[width=15cm]{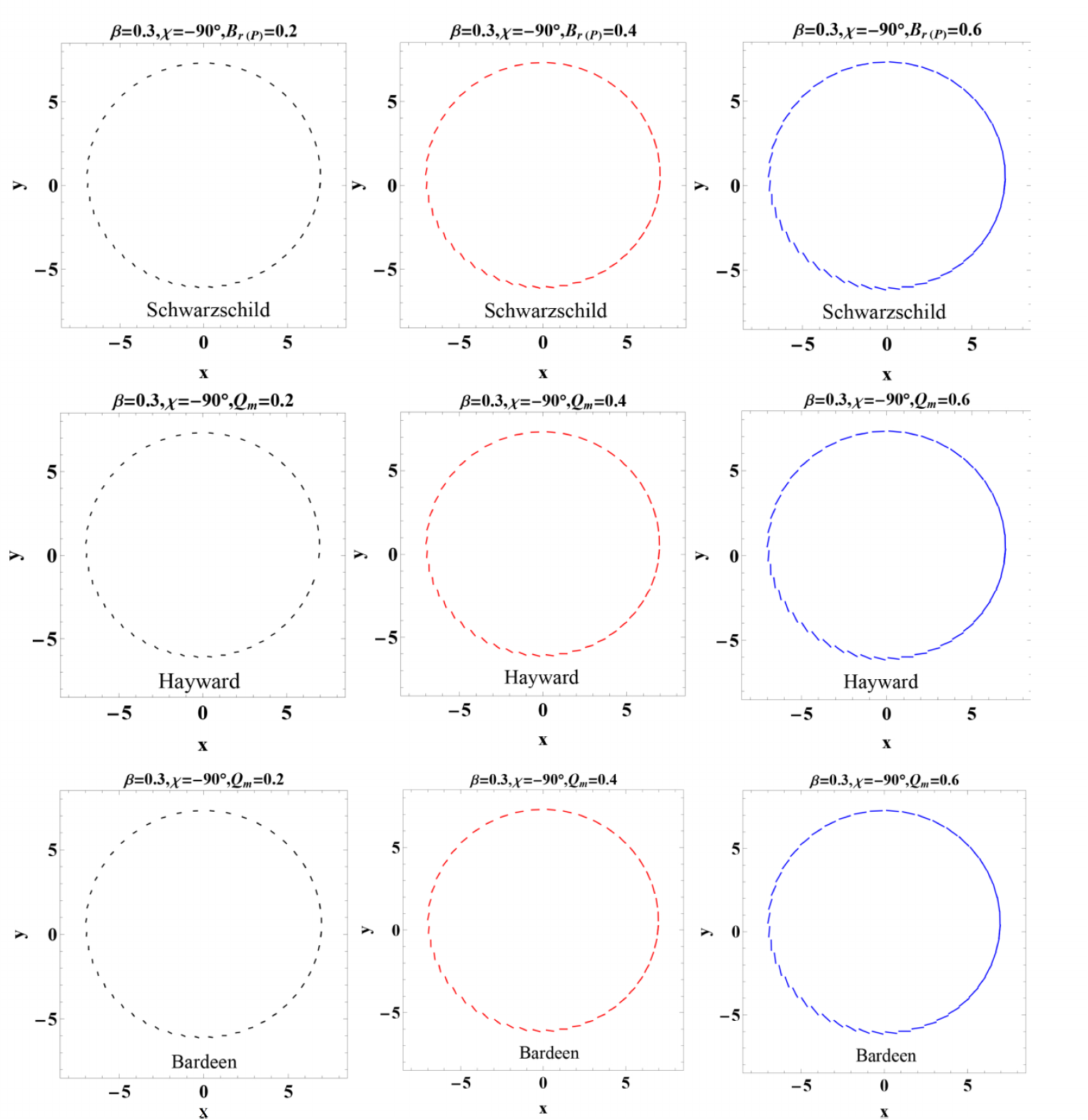}
\caption{Variation in the polarized image of a synchrotron emitting ring around a black hole with the magnetic field parameter. The first, second, and third lines correspond to the Schwarzschild, Hayward, and Bardeen black hole cases, respectively. Here, $M=1$, the observation inclination $\theta=20^{\circ}$, the emitting ring radius $R=6$, and the fluid velocity $\beta=0.3$.}
\label{rbp1}
\end{figure}

\begin{figure}
\centering
\includegraphics[width=14cm]{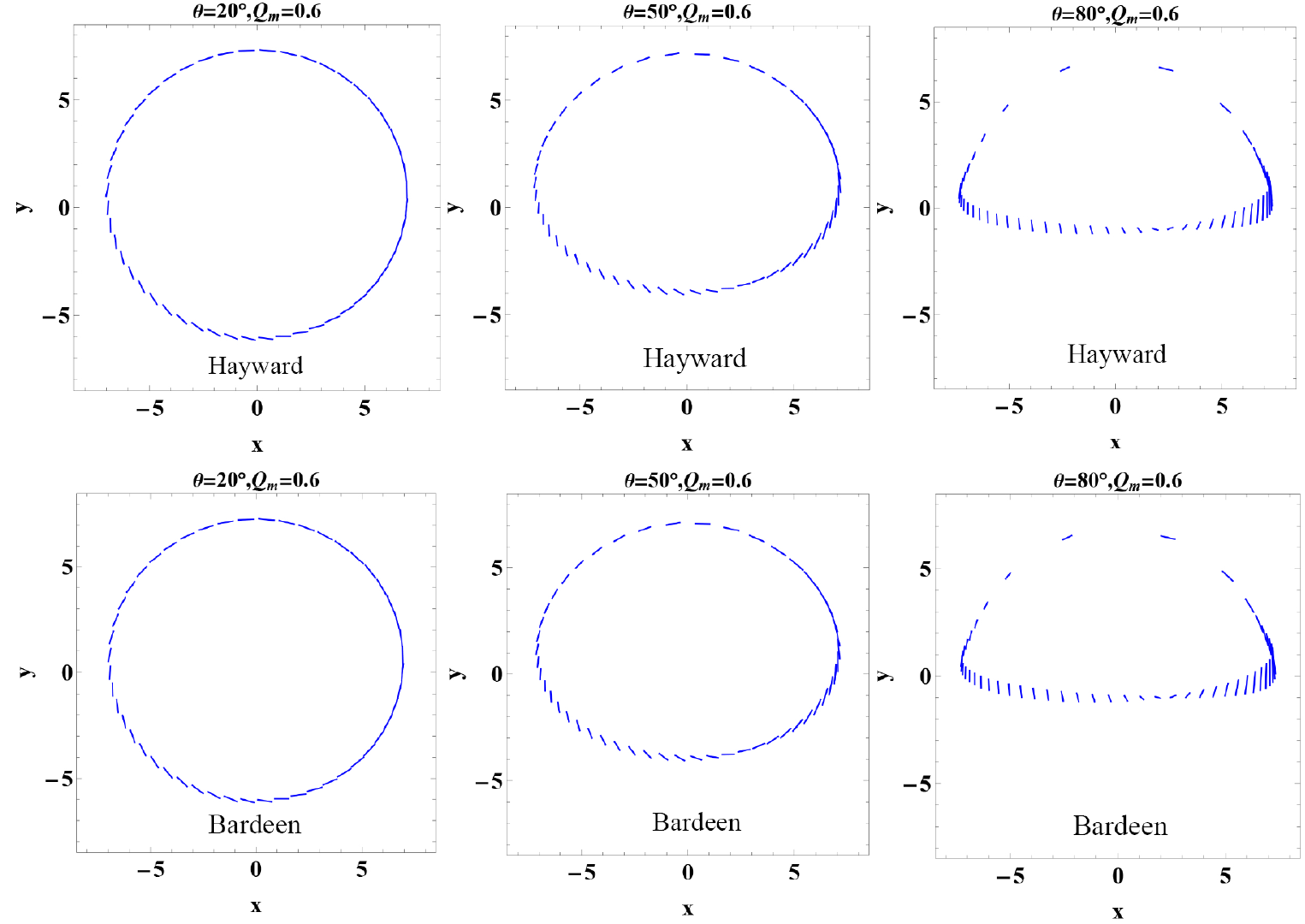}
\caption{Effect of the inclination angle $\theta$ on the polarized images of
a synchrotron emitting ring around a regular black hole for a fixed $Q_{m}=0.6$. The first and second lines correspond to Hayward and Bardeen black holes, respectively. Here, $M=1$, the ring radius $R=6$, and the fluid velocity $\beta=0.3$ and $\chi=-90^{\circ}$.}
\label{rbp3}
\end{figure}

\begin{figure}
\centering
\includegraphics[width=14cm]{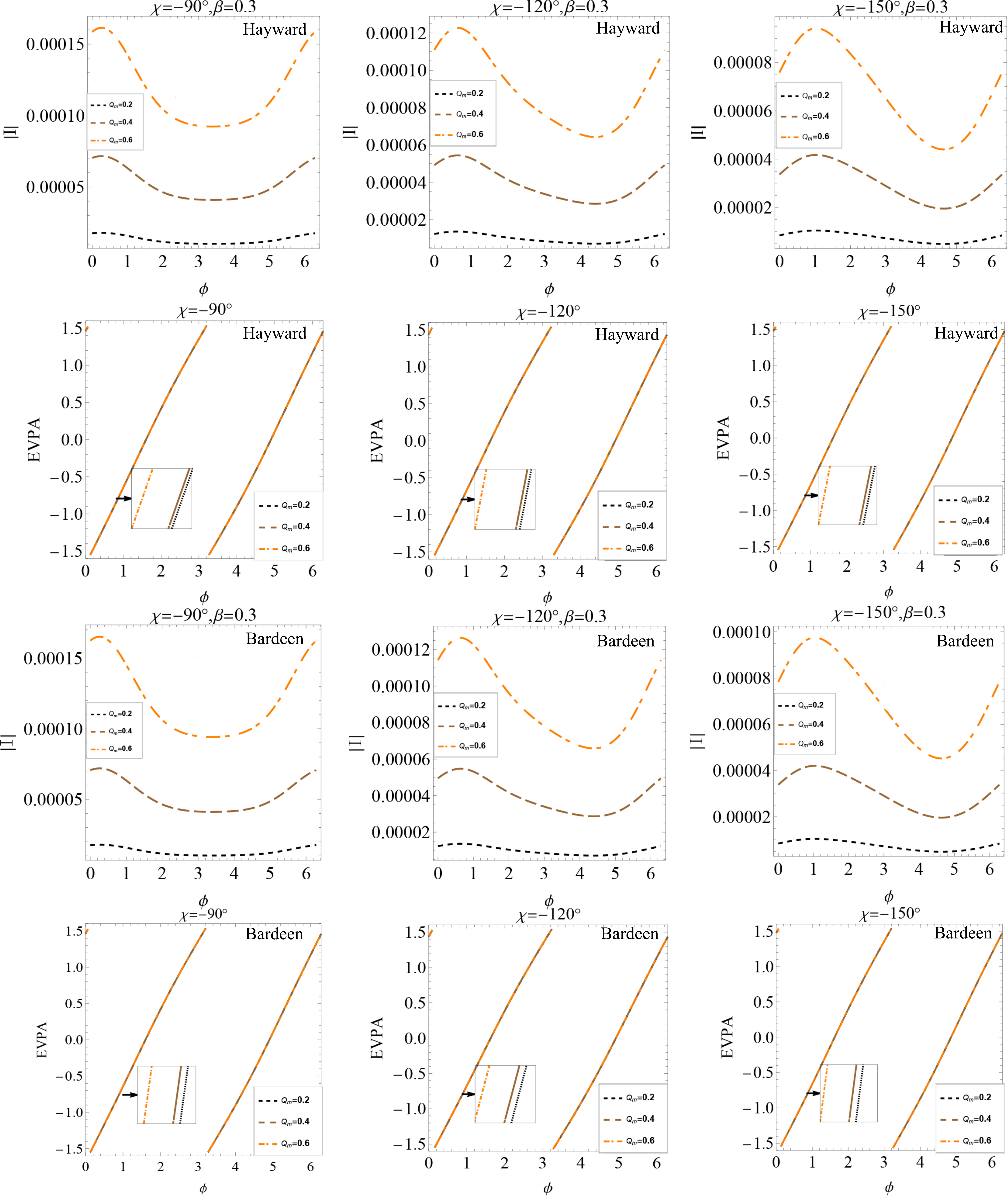}
\caption{Effect of the magnetic charge parameter $Q_{m}$ on the polarized intensity and the EVPA in the image plane for different values of $\chi$. The top two lines are for a Hayward black hole, and the bottom two lines are for a Bardeen black hole. Here, $M=1$, the ring radius $R=6$, the fluid velocity $\beta=0.3$, and the inclination angle $\theta=20^{\circ}$.}
\label{rb3}
\end{figure}

\begin{figure}
\centering
\includegraphics[width=14cm]{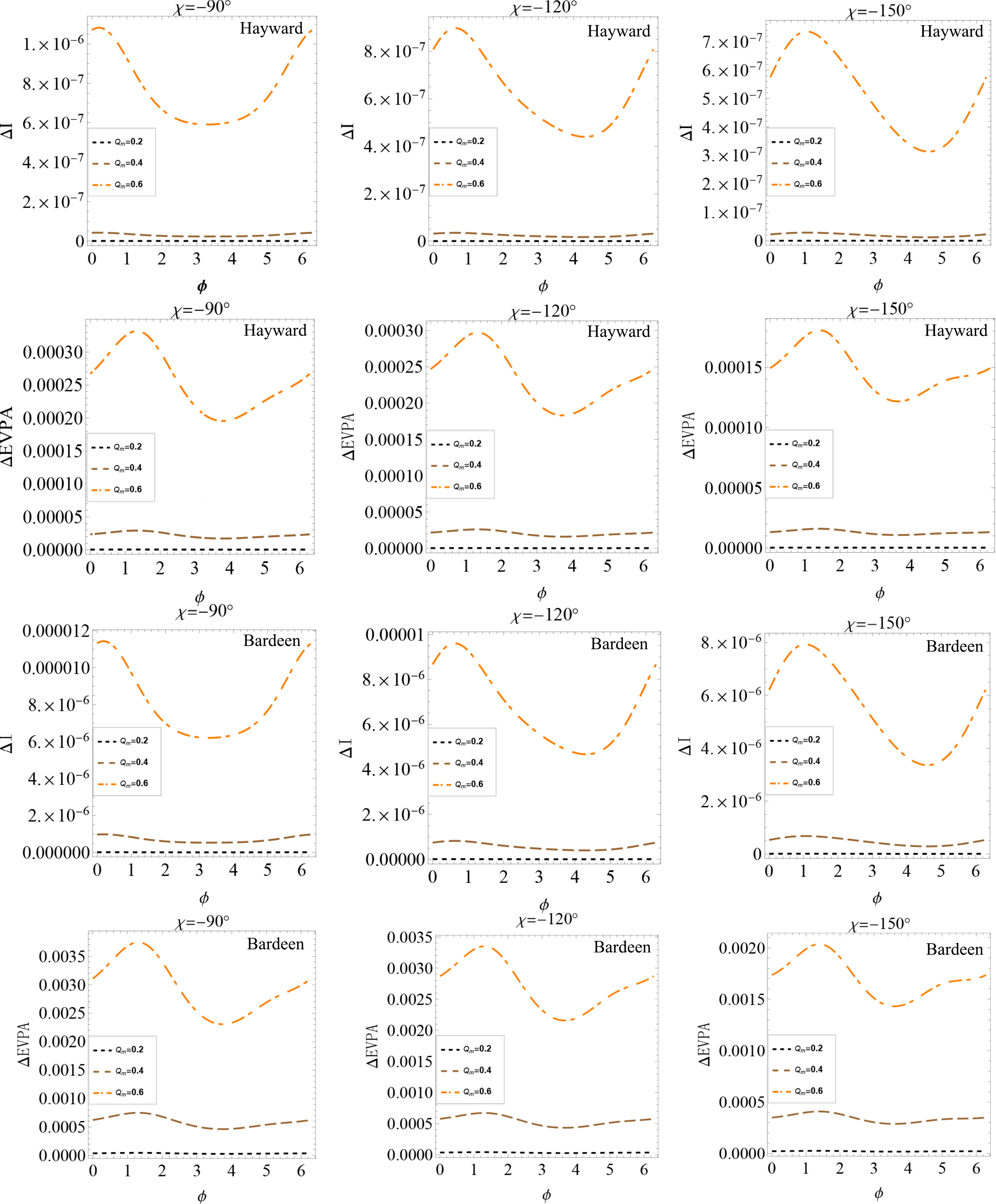}
\caption{Effect of the parameter $Q_{m}$ on the difference in the polarized intensity $\Delta I$ and in the electric vector position angle $\Delta EVPA $ for different values of $\chi$. The top two lines are for a Hayward black hole, and the bottom two lines are for a Bardeen black hole. Here, $M=1$, the ring radius $R=6$, the fluid velocity $\beta=0.3$, and the inclination angle $\theta=20^{\circ}$.}
\label{bh3}
\end{figure}

\begin{figure}
\centering
\includegraphics[width=14cm]{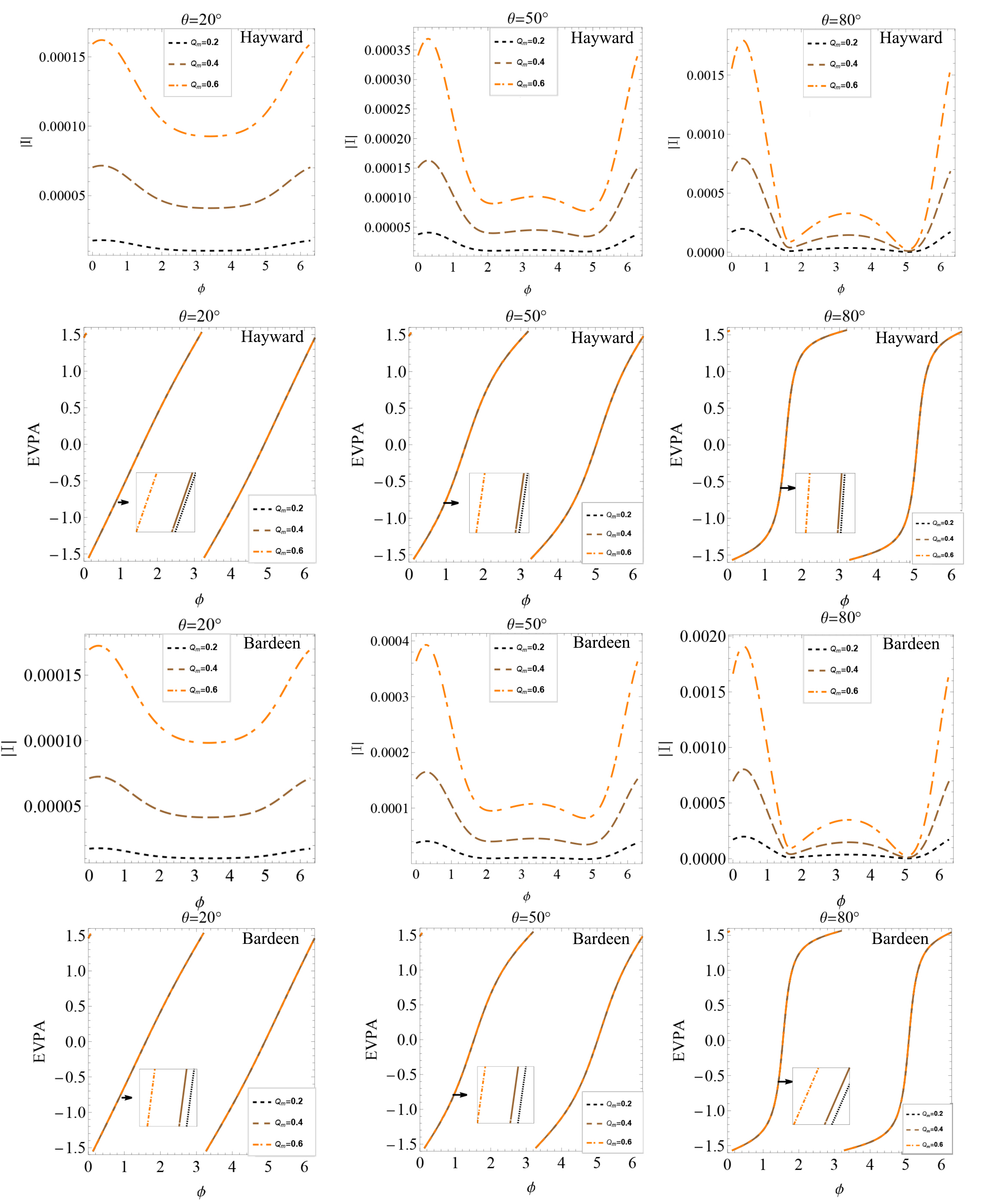}
\caption{Effects of the magnetic charge parameters $Q_{m}$ on the polarized intensity and the EVPA in the image plane for the different observation inclination angles $\theta$. The top two lines are for a Hayward black hole, and the bottom two lines are for a Bardeen black hole. Here, $M=1$, the ring radius $R=6$, the fluid velocity $\beta=0.3$, and the angle $\chi=-90^{\circ}$.}
\label{rb4}
\end{figure}

\begin{figure}
\centering
\includegraphics[width=14cm]{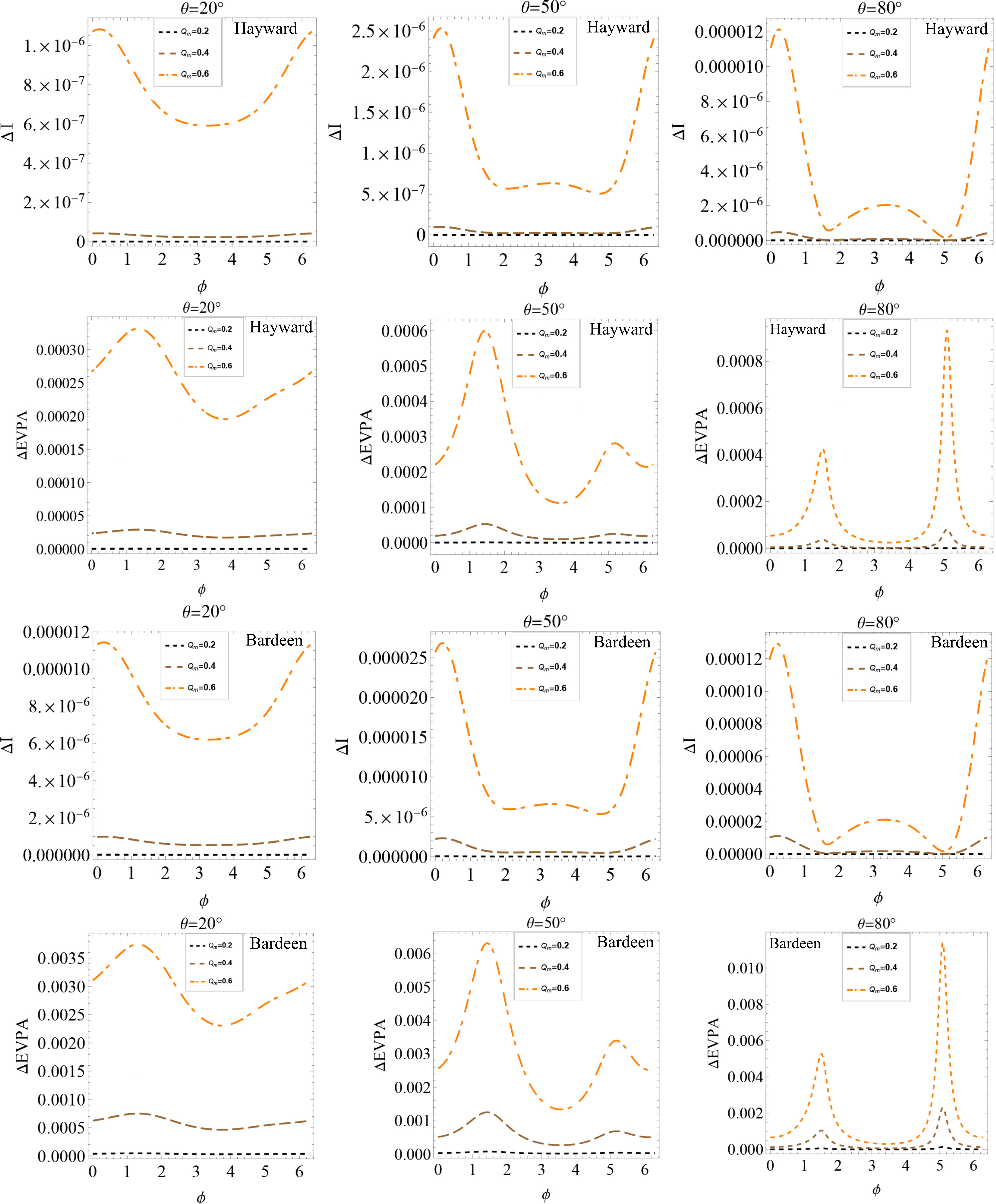}
\caption{Effects of the parameter $Q_{m}$ on the difference in the polarized intensity $\Delta I$ and of the electric vector position angle $\Delta EVPA $ for the different observation inclination angles $\theta$. The top two lines are for a Hayward black hole, and the bottom two lines are for a Bardeen black hole. Here, $M=1$, the ring radius $R=6$, the fluid velocity $\beta=0.3$, and the angle $\chi=-90^{\circ}$.}
\label{bh4}
\end{figure}

Figs. (\ref{rbp1})-(\ref{bh4}) present the polarization vector distribution in the image of the emitting ring (with radius $R=6$) around regular Hayward and Bardeen black holes, respectively.
The polarized intensity tick plots in Figs. (\ref{rbp1}) and (\ref{rbp3}) show that the properties of the observed polarized intensity in the image of the emitting ring for a fixed magnetic charge $Q_{m}$ and observer inclination angle $\theta$ are similar in the Hayward and Bardeen black hole spacetimes. Moreover, for a given external magnetic field, the polarization image features of the emitting ring in Schwarzschild and regular black hole spacetimes are qualitatively similar.

Figs. (\ref{rb3})-(\ref{bh4}) show that for fixed $\chi$ and $\beta$, the polarization intensity and EVPA change periodically with the angle coordinate $\phi$. Moreover, the polarization intensity increases with the magnetic charge $Q_m$ in the above two regular black hole spacetimes. EVPA also increases with $Q_m$, but the change amplitude is tiny. The changes in the quantities $\Delta I=I-I_{S}$ and $\Delta EVPA=EVPA-EVPA_{S}$ with the magnetic charge $Q_m$ are also presented in Figs. (\ref{bh3}) and (\ref{bh4}), where the subscript $S$ denotes the Schwarzschild black hole spacetime. These two quantities describe the difference between the polarization images of the emitting ring around a regular black hole and a Schwarzschild black hole. With increasing $Q_m$, the differences $\Delta I$ and $\Delta EVPA$ increase for the Hayward and Bardeen black holes. However, the values of $\Delta I$ and $\Delta EVPA$ are smaller in the Hayward black hole spacetime than in the Bardeen black hole, which is due to the metric function for the Hayward black hole being closer to that for the Schwarzschild black hole since $f_{\rm Hayward}\approx 1-\frac{2M}{r}+\frac{16Q^6_m}{M^2r^4}+\mathcal{O}(Q^{10}_m)$ and $f_{\rm Bardeen}\approx 1-\frac{2M}{r}+\frac{12Q^4_m}{Mr^3}+\mathcal{O}(Q^6_m)$, which means that the effect of $Q_m$ arising from a metric function is smaller in the Hayward black hole spacetime.
For different observer inclination angles $\theta$, Figs. (\ref{rb4})-(\ref{bh4}) also show that the polarization intensity and EVPA change periodically with the angle coordinate $\phi$ in both regular black hole spacetimes. With increasing $Q_m$, the polarization intensity, EVPA, and differences $\Delta I$ and $\Delta EVPA$ increase. Moreover, the values of $\Delta I$ and $\Delta EVPA$ are smaller in Hayward black hole spacetime than in a Bardeen black hole, as with the cases with different fluid velocity angles $\chi$.

\begin{figure}
\includegraphics[width=14cm]{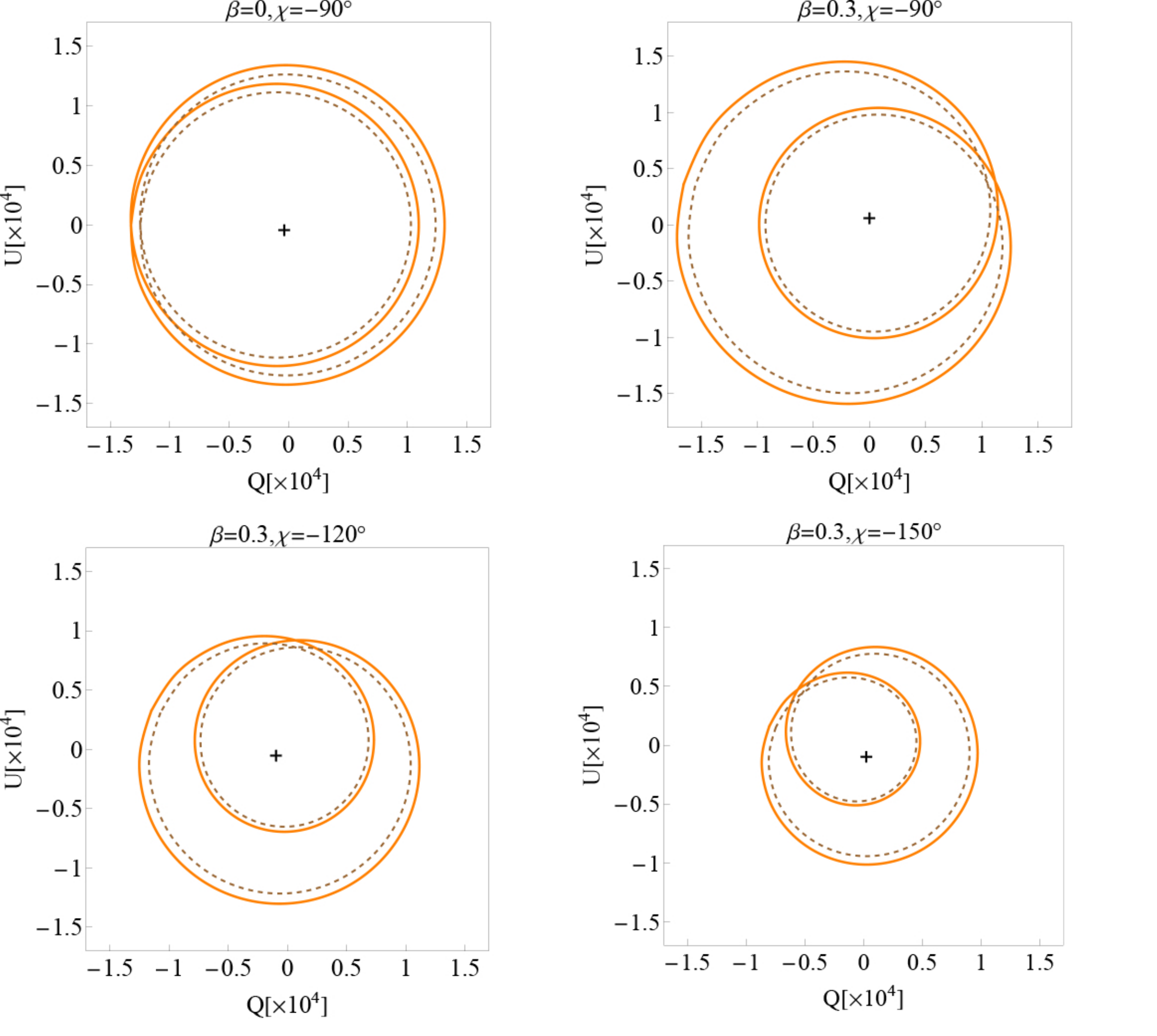}
\caption{$Q-U$ loops diagram of the polarization vector in the image of the emitting ring around a regular black hole for different fluid velocity angles $\chi$. The brown dotted line corresponds to a Hayward black hole, and the orange solid line corresponds to a Bardeen black hole. Black crosshairs indicate the
origin of each plot. Here, $M=1$, $R=6$, $\theta=20^{\circ}$, $\beta=0.3$, and $Q_{m}=0.6$.}
\label{uq1}
\end{figure}

\begin{figure}
\includegraphics[width=16cm]{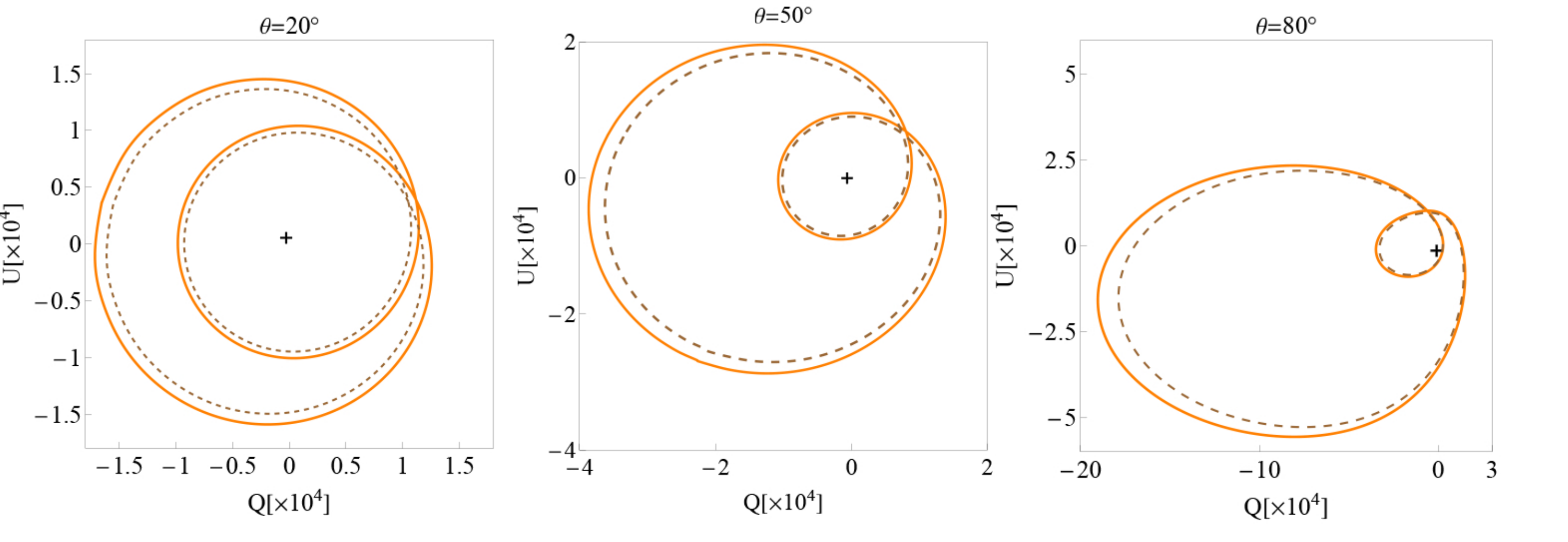}
\caption{$Q-U$ loops diagram of the polarization vector in the image of the emitting ring around a regular black hole for different observing inclination angles $\theta$. The brown dotted line corresponds to a Hayward black hole, and the orange solid line corresponds to a Bardeen black hole. Black crosshairs indicate the
origin of each plot. Here, $M=1$, $R=6$, $\chi=-90^{\circ}$, $\beta=0.3$, and $Q_{m}=0.6$.}
\label{uq2}
\end{figure}

The $Q-U$ loop patterns of the polarization vector are studied in the image of the emitting ring around a regular black hole and describe the continuous variability of the polarization vector in the image of the emitting ring around a regular black hole. As with the usual static black holes, two loops surround the origin in the $Q-U$ plane. Figs. (\ref{uq1}) and (\ref{uq2}) show that the $Q-U$ loops are similar in shape in the Hayward and Bardeen black hole spacetimes but slightly smaller for the Hayward black hole. With increasing fluid direction angle $\chi$, the inner and outer rings shrink. With increasing inclination angle $\theta$, the outer loop gradually expands and the inner loop dramatically shrinks such that the $Q-U$ loops gradually change from circular to irregular-shaped. With increasing $Q_m$, the loops change very slightly.
This information stored in the polarization image could help understand regular black holes and gravity-coupled nonlinear electrodynamics.

\section{Summary}

This study investigated the polarized images of the emitting ring around regular Hayward and Bardeen black holes. The results show that the polarization images and polarization characteristics of the emitting rings are similar in these two regular black holes. The dependence of the polarization image of the emitting ring on the fluid velocity and observer inclination angle is similar to that in the usual static black hole spacetimes. With the increase in magnetic charge parameter $Q_{m}$, the polarization intensity and EVPA of each point in the image plane increase in Hayward and Bardeen black hole spacetimes. The difference values $\Delta I$ and $\Delta EVPA$ between regular and the Schwarzschild black holes with the same magnetic field increase with $Q_{m}$. However, these difference values are smaller in Hayward black hole spacetimes than in Bardeen black hole spacetimes.

Moreover, the $Q-U$ loop patterns of the polarization vector are studied in the image of the emitting ring around a regular black hole. The $Q-U$ loops are similar in shape in the Hayward and Bardeen black hole spacetimes but slightly smaller in the Hayward black hole. With increasing fluid direction angle $\chi$, the inner and the outer rings shrink. With increasing inclination angle $\theta$, the outer loop gradually expands and the inner loop dramatically shrinks, making the shapes of $Q-U$ loops more irregular. With increasing magnetic charge parameter $Q_m$, the loops change very slightly. This information stored in the polarization image around Hayward and Bardeen black holes could help understand regular black holes and the gravity coupled to nonlinear electrodynamics.

\section{\bf Acknowledgments}

This work was  supported by the National Natural Science
Foundation of China under Grant No.11875026,  12035005 and 2020YFC2201400.

\end{document}